\documentstyle[12pt,psfig,axodraw,a4]{article} 
\def\bild#1#2{    
        \vspace*{-5mm}
        \begin{center}
        \begin{math}
        \epsfxsize#2cm
        \epsffile{#1}
        \end{math}
        \end{center}
        }
\newcommand{\vs}{\vspace{-0.25cm}}

\begin{document} 

\hfill TUM/T39-01-19

\begin{center}
\Large{\bf Nuclear mean field from chiral pion-nucleon dynamics}\footnote{
Work supported in part by BMBF, GSI and DFG.} 

\bigskip 

\bigskip

N. Kaiser$\,^a$, S. Fritsch$\,^a$ and W. Weise$\,^{a,b}$\\

\bigskip

{\small
$^a$\,Physik Department, Technische Universit\"{a}t M\"{u}nchen, D-85747
Garching, Germany\\

\smallskip

$^b$\, ECT$^*$, I-38050 Villazzano (Trento), Italy\\

\smallskip

{\it email: nkaiser@physik.tu-muenchen.de}}

\end{center}

\bigskip

\begin{abstract}
Using the two-loop approximation of chiral perturbation theory, we calculate 
the momentum and density dependent single particle potential of nucleons in 
isospin-symmetric nuclear matter. The contributions from one- and two-pion 
exchange diagrams give rise to a potential depth for a nucleon at rest of 
$U(0,k_{f0}) = -53.2\,$MeV at saturation density. The momentum dependence of 
the real part of the single particle potential $U(p,k_{f0})$ is non-monotonic
and can be translated into a mean effective nucleon mass of $\bar M^* 
\simeq 0.8 M$. The imaginary part of the single particle potential $W(p,k_f)$
is generated to that 
order entirely by iterated one-pion exchange. The resulting half width of a
nucleon hole-state at the bottom of the Fermi sea comes out as $W(0,k_{f0})=
29.7\,$MeV. The basic theorems of Hugenholtz-Van-Hove and Luttinger are
satisfied in our perturbative two-loop calculation of the nuclear mean field. 
\end{abstract}

\bigskip
PACS: 12.38.Bx, 21.65.+f\\
Keywords: Effective field theory at finite density, Real and imaginary part of
          the single particle potential in nuclear matter, effective nucleon
          mass. 

\vskip 1.5cm

\section{Introduction and summary}
The shell model and the optical model have always played a central role in the
description of 
nuclear structure and nuclear reactions. The knowledge of single particle 
properties in dense nuclear matter is therefore of basic importance for the 
understanding of both nuclear structure and nuclear dynamics. Quantities like 
the nuclear mean field (or equivalently the single particle potential), the 
nucleon mean free path, the effective mass, etc., enter in the description of 
static as well as dynamic properties of nuclei (for a review see 
ref.\cite{mahaux}). Considerable progress has been made during the past  
decades in calculating these single particle quantities from first principles,
based on non-relativistic Brueckner theory \cite{jeuk,grange} and relativistic 
many-body theory \cite{malfliet,rolf}. Despite the widely different mechanisms 
which build up the nuclear mean field in these two theoretical approaches, the 
depth of the single particle potential comes out at approximately the same 
value when both calculations are restricted to first order in the
renormalized in-medium NN-interaction (G-matrix). Other properties, such as the
momentum dependence of the nuclear mean field are also qualitatively similar 
in those two approaches. 

In a recent work \cite{nucmat}, we have used chiral perturbation theory for a  
systematic treatment of the nuclear matter many-body problem. In this 
calculation the contributions to the energy per particle, $\bar E(k_f)$,
originate exclusively from one- and two-pion exchange between nucleons and they
are ordered in powers of the Fermi momentum $k_f$ (modulo functions of
$k_f/m_\pi$ with $m_\pi$ denoting the pion mass). It has been demonstrated in 
ref.\cite{nucmat} that the empirical saturation point and the nuclear matter
compressibility $K\simeq 250\,$MeV can be well reproduced at order ${\cal
O}(k_f^5)$ in the chiral 
expansion with just one single momentum cut-off scale of $\Lambda \simeq
0.65\,$GeV which parametrizes all necessary short range dynamics. Most
surprisingly, the prediction for the asymmetry energy, $A_0=33.8\,$MeV, is in
very good agreement with its empirical value. Furthermore, as a nontrivial fact
pure neutron matter is predicted to be unbound and the corresponding equation
of state agrees roughly with that of sophisticated many body calculations for
low neutron densities $\rho_n \leq 0.25\,$fm$^{-3}$. 

Given the fact that the bulk properties of nuclear matter can be well described
by chiral pion-nucleon dynamics treated up to three loop order it is natural to
consider in a next step quantities which characterize in more detail the 
behavior of a single nucleon moving in the dense nuclear matter environment. 
It is the purpose of this work to calculate, using the same framework as in
ref.\cite{nucmat}, the momentum and
density dependent (complex-valued) single particle potential of nucleons in 
isospin-symmetric nuclear matter. We will present here analytical expressions 
for both the real part $U(p,k_f)$ and the imaginary part $W(p,k_f)$ of the 
nuclear mean field generated by chiral one- and two-pion exchange. Our
results can be summarized as follows:
\begin{itemize}
\item[i)] At equilibrium nuclear matter density chiral one- and two-pion 
exchange generate an attractive nuclear mean field for nucleons at rest, with
a depth of $U(0,k_{f0})=-53.2\,$MeV. This value is in very good agreement with 
the depth of the empirical optical model potential $U_0 \simeq -52\,$MeV 
deduced by extrapolation from elastic nucleon-nucleus scattering data 
\cite{phenpot,hodgson}. In fact, the average nuclear potential employed in 
shell model calculations of nuclei has approximately the same depth
\cite{bohr}.   

\item[ii)] The momentum dependence of the real part of the single particle
potential $U(p,k_{f0})$ is non-monotonic in the momentum interval $0\leq p\leq
k_{f0}$. Its maximal increase of $\Delta U= 8.8\,$MeV can be translated into 
an average effective nucleon mass of $\bar M^* = 0.82 M$. This is 
comparable to the empirical value of the effective nucleon mass in nuclear 
matter, $M^*_{emp} = (0.7 - 0.8) M$, derived from experimental data in the
framework of non-relativistic shell or optical models 
\cite{horen,jaminon,brack}.  

\item[iii)] To the order considered here the imaginary part of the single 
particle potential $W(p,k_f)$ is entirely generated by iterated one-pion
exchange. For a nucleon hole-state at the bottom of the Fermi sea, one finds a 
half width $\Gamma_{hole}/2= W(0,k_{f0}) =29.7\,$MeV.  This value is not far 
from results obtained in Brueckner-Hartree-Fock calculations based on realistic
NN-forces, where $W(0,k_{f0}) \simeq 40 \,$MeV \cite{grange}.

\item[iv)] The Hugenholtz-Van-Hove theorem \cite{vanhove} which relates the 
total single particle energy at the Fermi surface (i.e. the Fermi energy) to
the nuclear matter equation of state is satisfied in our two-loop chiral
perturbation theory calculation. Luttinger's theorem \cite{luttinger} does also
hold since the calculated imaginary part, $W(p,k_f)\sim (k_f-p)^2 $, vanishes
quadratically near the Fermi surface. The associated curvature coefficient
$C(k_f)$ grows with density approximately as $\rho^{2/3}$. 
\end{itemize}

As a combined conclusion of ref.\cite{nucmat} and this work one can thus state
that perturbative chiral pion nucleon dynamics gives realistic nuclear binding
and saturation as well as realistic (in-medium) single particle properties.   

\section{Single particle potential of nucleons in nuclear matter}
The nuclear mean field (or single particle potential) is generally defined
through the momentum and density dependent nucleon self energy in (isospin 
symmetric) nuclear matter. In a diagrammatic calculation this quantity is
derived from the in-medium nucleon propagator taking into account interactions
among nucleons up to a certain order as well as Pauli-blocking effects. On the 
other hand, the ground state energy density of nuclear matter
is represented diagrammatically by analogous closed vacuum graphs which 
result from closing the nucleon line of a self energy diagram . As outlined in 
section 2 of ref.\cite{nucmat} the diagrammatic calculation of the ground state
energy density can be organized in the number of so-called medium insertions. 
The latter is a technical notation for the difference between the vacuum and 
the bare in-medium nucleon propagator. Let us consider the structure of the 
energy density according to such an ordering scheme \cite{nucmat}. It
consists of a sum of convolution integrals of the form,
\begin{eqnarray} {\cal E}[d]&=&\int d^3 p_1 \, {\cal K}_1 \,d(\vec p_1)+ 
\int d^3 p_1d^3p_2\,{\cal K}_2\,d(\vec p_1)d(\vec p_2) + \int d^3 p_1d^3 p_2
d^3p_3\,{\cal K}_3 \,d(\vec p_1) d(\vec p_2) d(\vec p_3) \nonumber \\ &&
+ \int d^3 p_1d^3 p_2d^3p_3 d^3p_4\,{\cal K}_4 \,d(\vec p_1) d(\vec p_2) d(\vec
p_3)d(\vec p_4) \,. \end{eqnarray} 
The one-body kernel ${\cal K}_1= 4T_k(|\vec p_1|)$ is four times the 
relativistically improved kinetic energy (see eq.(3)). ${\cal K}_{2,3}$ 
are two- and three-body kernels related to contributions of closed diagrams 
with two and three medium insertions. In three-loop approximation the four-body
kernel ${\cal K}_4$ is proportional to $\delta^3( \vec p_1+\vec p_2+\vec p_3+
\vec p_4)$ and purely imaginary. The quantity $d(\vec p_j)$ denotes the density
of states in momentum space. Inserting the density of states of a filled Fermi
sea, $d(\vec p_j)= (2\pi)^{-3} \,\theta(k_f-|\vec p_j|)$, into eq.(1) one gets
the energy density of nuclear matter, $\rho \bar E(k_f)$, with the nucleon
density $\rho= 2k_f^3/3\pi^2$. The single particle potential can now be
directly constructed from the energy density functional eq.(1) by adding a test
nucleon of fixed momentum $\vec p$ to the filled Fermi sea. 
This situation is described by the density of states $d(\vec p_j)= 
(2\pi)^{-3} \, \theta(k_f-|\vec p_j|)+ \eta  \,\delta^3(\vec p -\vec p_j)$
with the infinitesimal parameter $\eta$ to be interpreted as $(\pm)$ the 
inverse (infinite) volume. The plus sign applies for a particle ($|\vec p\,| >
k_f)$ and the minus sign for a hole $(|\vec p\,| < k_f)$. Inserting this
density of states into eq.(1) leads to 
\begin{equation}  {\cal E}= \rho \, \bar E(k_f) + 4 \eta \, \Big\{ T_k(p)+ 
U(p,k_f) + i \,W(p,k_f) \Big\} \,, \end{equation}
with
\begin{equation} T_k(p) = {p^2 \over 2M}-{p^4 \over 8M^3}\,,\end{equation}
the relativistically improved kinetic energy. The factor 4 in eq.(2) simply 
counts the spin- and isospin multiplicity of a nucleon. The real and imaginary
parts of the single particle potential have, according to eq.(1), a 
decomposition into two-, three- and four-body contributions,  
\begin{equation}
U(p,k_f)=U_2(p,k_f)+U_3(p,k_f) \,, \qquad W(p,k_f)=W_2(p,k_f)+W_3(p,k_f)+
W_4(p,k_f)  \,,\end{equation}
where the index on each term refers to the corresponding kernel ${\cal K}_n$ in
eq.(1). The energy per particle $\bar E(k_f)$ in symmetric nuclear matter can 
also be reconstructed from these components of the single particle potential by
integrating over a Fermi-sphere of radius $k_f$, 
\begin{equation} \bar E(k_f) = {3\over k_f^3} \int_0^{k_f} dp \, p^2 \bigg[
T_k(p) + {1\over 2} U_2(p,k_f) +{1\over 3} U_3(p,k_f)\bigg] \,. \end{equation}
Note the weighting factors $1/2$ and $1/3$. These take care of combinatoric
factors and the fact that three topologically equivalent closed vacuum diagrams
with three medium insertions are generated by closing the nucleon line of a
self energy diagram. Since this closing procedure must lead to a real energy
per particle $\bar E(k_f)$ the components of the imaginary part of the single
particle potential have to fulfill the zero sum rule
\begin{equation}{3\over k_f^3}\int_0^{k_f} dp \, p^2 \bigg[{1\over 2}W_2(p,k_f)
+{1\over 3} W_3(p,k_f)+ {1\over 4} W_4(p,k_f)\bigg] = 0\,. \end{equation}
A further constraint on the single particle potential is given by the 
Hugenholtz-van-Hove theorem \cite{vanhove}. It states that the total single 
particle energy at the Fermi surface, $p=k_f$, (i.e. the Fermi energy) is equal
to the chemical potential. According to a general thermodynamic relation this
chemical potential is the derivative of the energy density $\rho \bar E(k_f)$ 
with respect to the particle density $\rho$. The Hugenholtz-van-Hove theorem 
reads explicitly,   
\begin{equation} T_k(k_f)+ U(k_f,k_f) = \bar E(k_f) + {k_f \over 3}\,{\partial 
\bar E(k_f) \over \partial k_f} \,, \qquad W(k_f,k_f)=0\,. \end{equation}
Note that the derivative term $\partial \bar E(k_f) / \partial k_f$  
vanishes at nuclear matter saturation density and therefore the Fermi energy 
of equilibrated nuclear matter is equal to the negative average binding energy
per particle, $\bar E(k_{f0})$. Eqs.(5,6,7) serve as important numerical and 
analytical checks on our chiral perturbation theory calculation to be presented
in the next two sections.  

\section{Real part of the single particle potential}

\begin{center}

\SetWidth{2.5}
  \begin{picture}(400,100)

\ArrowArc(60,50)(40,0,180)
\ArrowArc(60,50)(40,180,360)
\DashLine(20,50)(100,50){6}
\Vertex(20,50){4}
\Vertex(100,50){4}

\CArc(180,50)(40,90,270)
\ArrowLine(180,10)(180,90)
\DashLine(180,90)(240,90){6}
\Vertex(180,90){4}
\Vertex(240,90){4}
\CArc(240,50)(40,-90,90)
\ArrowLine(240,10)(240,90)
\DashLine(180,10)(240,10){6}
\Vertex(180,10){4}
\Vertex(240,10){4}

\ArrowArc(350,50)(40,0,180)
\ArrowArc(350,50)(40,180,360)
\DashLine(378.3,78.3)(321.7,21.7){6}
\DashLine(378.3,21.7)(321.7,78.3){6}
\Vertex(378.3,78.3){4}
\Vertex(321.7,21.7){4}
\Vertex(378.3,21.7){4}
\Vertex(321.7,78.3){4}

\end{picture}
\end{center}
{\it Fig.\,1: One-pion exchange Fock-diagram and iterated one-pion exchange 
Hartree- and Fock-diagrams. The combinatoric factor of these diagrams are 1/2,
1/4 and 1/4, in the order shown. Each nucleon propagator consists of a vacuum
part and a medium insertion.}

\bigskip

In this section we present analytical results for the real part of the single
particle potential $U(p,k_f)$ as given by chiral one- and two-pion exchange. 
The closed vacuum diagrams related to one-pion exchange (Fock-diagram) and 
iterated one-pion exchange (Hartree- and Fock-diagrams) are shown in Fig.\,1. 
Each nucleon propagator consists of a vacuum part and a medium insertion (for
details see eq.(3) in ref.\cite{nucmat}). Self energy diagrams are obtained
from these closed graphs by opening one nucleon line with a medium
insertion. In order to keep their number small and in order to avoid
repetitions we associate the contributions to the (real) nuclear mean field
$U(p,k_f)$ with the closed vacuum diagram before opening of a nucleon line. The
index $n$ on $U_n(p,k_f)$ (see eq.(4)) then indicates the number of medium
insertions of the closed vacuum diagram before line opening. From a technical
point of view, the calculation of a contribution $U_n(p,k_f)$ (or $W_n(p,k_f)$)
involves an integral over the product of $n-1$ Fermi spheres of radius $k_f$. 

We start with the $1\pi$-exchange Fock-diagram in Fig.\,1 with two medium
insertions. With inclusion of the relativistic $1/M^2$-correction one finds a 
density and momentum dependent real two-body potential of the form,
\begin{eqnarray} U_2(p,k_f) &=& {3g_A^2 m_\pi^3 \over (4\pi f_\pi)^2} \bigg\{
{2u^3 \over 3}-u +\arctan(u+x)+\arctan(u-x) +{x^2-u^2-1 \over 4x}\ln{1+(u+x)^2
\over 1+(u-x)^2} \nonumber \\ && +{m_\pi^2 \over 16 M^2} \bigg[ 4u\Big( 1-{u^2
\over 3} -{4u^4 \over 5}+3x^2 -{4\over 3}u^2x^2 \Big) -(2+8x^2) \Big[ 
\arctan(u+x)\nonumber \\ && +\arctan(u-x)\Big]  +{1\over x}(u^2-x^2)(u^2+3x^2) 
\ln{1+(u+x)^2\over 1+(u-x)^2} \bigg] \bigg\}\,, \end{eqnarray}
where we have introduced the abbreviations $u=k_f/m_\pi$ and $x=p/m_\pi$. 
As in ref.\cite{nucmat}, we choose the value $g_A=1.3$ for the nucleon axial 
vector coupling constant. $f_\pi = 92.4\,$MeV denotes the weak pion decay
constant and $m_\pi=135\,$MeV stands for the (neutral) pion mass. The iterated
$1\pi$-exchange Hartree-diagram (second graph in Fig.\,1) with two medium 
insertions gives rise to a real part of the single particle potential which 
reads
\begin{eqnarray} U_2(p,k_f) &=& {g_A^4Mm_\pi^4 \over (4\pi)^3 f_\pi^4} \bigg\{
\Big( {9\over 2}+3u^2+{2u^3 \over x}-x^2 \Big) \arctan(u+x) + \Big( {9\over 2}
+3u^2  -{2u^3 \over x}-x^2 \Big) \nonumber \\ && \times \arctan(u-x) - {11 u
\over 2} +{1 \over 8x} (15x^2-15u^2-7) \ln{1+(u+x)^2 \over 1+(u-x)^2}
\bigg\}\,. \end{eqnarray} 
In this expression we have omitted the contribution of a linear divergence 
proportional to the cut-off $\Lambda$ (see ref.\cite{nucmat}). All such
powerlike terms in $\Lambda$ are collected in eq.(17). The iterated $1\pi
$-exchange Fock-diagram (third graph in Fig.\,1)  with two medium insertions
gives rise to a further contribution to the two-body potential $U_2(p,k_f)$ of
the form,      
\begin{eqnarray} U_2(p,k_f) &=& {g_A^4Mm_\pi^4 \over (4\pi)^3 f_\pi^4} \bigg\{
u^3 +\bigg[ \int_0^{(u-x)/2}\!\! d\xi\, 2\xi+ \int_{(u-x)/2}^{(u+x)/2}\!\! d\xi
{1\over 4x}(u^2-(2\xi-x)^2 ) \bigg]\nonumber \\ && \times  {3\over 1+2\xi^2} 
\Big[(1+8\xi^2 + 8\xi^4) \arctan\xi-(1+4\xi^2) \arctan 2\xi \Big] \bigg\} \,,
\end{eqnarray}  
where we have again transferred a term linear in the cut-off $\Lambda$ to
eq.(17). The notation in eq.(10) is to be understood such that the factor in 
the second line belongs to both integrals. Next, we consider the iterated 
$1\pi$-exchange Hartree-diagram (second graph in Fig.\,1) with three medium 
insertions. In this case  one gets three different contributions to the
three-body potential $U_3(p,k_f)$ corresponding to the three possibilities of
opening a nucleon line. Altogether they read,   
\begin{eqnarray} U_3(p,k_f)&=&{6g_A^4M m_\pi^4\over (4\pi f_\pi)^4}\int_{-1}^1 
\!dy\bigg\{\bigg[2uxy+(u^2-x^2y^2)\ln{u+xy\over u-xy}\bigg]\bigg[{2s^2+s^4\over
2( 1+s^2)}-\ln(1+s^2)\bigg] \nonumber \\ && + \int_{-xy}^{s-xy}\!\!d\xi\,
\bigg[ 2u\xi +(u^2-\xi^2)\ln{u+\xi \over u-\xi} \bigg]\,{(xy
+\xi)^5 \over [1+ (xy +\xi)^2]^2} \nonumber \\ &&+ \int_0^u \!\!d\xi \, {\xi^2 
\over x} \ln {|x+\xi y|\over |x-\xi y|}\, \bigg[{2\sigma^2 +\sigma^4\over
1+\sigma^2}- 2\ln(1+\sigma^2)\bigg] \bigg\} \,, \end{eqnarray}
with the auxiliary functions, 
\begin{equation}s= xy+\sqrt{u^2-x^2+x^2y^2}\,, \qquad \sigma= \xi y+\sqrt{u^2-
\xi^2+\xi^2y^2}\,.  \end{equation}
Note that the expressions in eq.(11) originate from a six-dimensional principal
value integral over the product of two Fermi spheres of radius $k_f$. The 
actual integrands  have simple poles located on a five-dimensional quadratic
hypersurface. The  same features hold for the iterated $1\pi$-exchange
Fock-diagram (third graph in Fig.\,1) with three medium insertions. Its
contribution to the three-body potential $U_3(p,k_f)$ reads
\begin{eqnarray} U_3(p,k_f) &=&{3g_A^4M m_\pi^4 \over (4\pi f_\pi)^4} \bigg\{
{G^2(x) \over 8x^2} +\int_0^u \!\! d\xi \, G(\xi) \bigg[ 1 +{\xi^2-x^2-1
\over 4 x \xi }\ln{1+(x+\xi)^2 \over 1+(x-\xi)^2} \bigg] \nonumber \\ && + 
\int_{-1}^1\!dy  \bigg[ \int_{-1}^1\!dz {yz \,\theta(y^2+z^2-1) \over 4|yz|
\sqrt{y^2+ z^2-1}}\Big[s^2-\ln(1+s^2) \Big] \Big[\ln(1+t^2)-t^2\Big] \nonumber
\\ && + \int_0^u \!d\xi \, {\xi^2\over x} \Big[ \ln(1+\sigma^2)-\sigma^2\Big]
\bigg( \ln {|x+\xi y|\over |x-\xi y|} \nonumber \\ &&  +{1\over R} \ln{[x
R+(x^2-\xi^2-1)y \xi]^2 \over [1+(x+\xi)^2][1+(x-\xi)^2)] |x^2-\xi^2 y^2|}
\bigg) \bigg] \bigg\}\,, \end{eqnarray}
where we have introduced some new auxiliary functions,
\begin{equation} G(x) = u(1+u^2+x^2) -{1\over 4x}\big[1+(u+x)^2\big] \big[1+
(u-x)^2\big] \ln{1+(u+x)^2\over 1+(u-x)^2 } \,, \end{equation} 
\begin{equation} t= xz+\sqrt{u^2-x^2+x^2z^2}\,,\qquad R= \sqrt{(1+x^2-\xi^2)^2
+4\xi^2(1-y^2)}\,. \ \end{equation}
Note that all terms from iterated $1\pi$-exchange are proportional to the
(large) nucleon mass, $M=939\,$MeV. The additional diagrams of irreducible 
$2\pi$-exchange (not shown here, but see Fig.\,4 in ref.\cite{nucmat}) with two
medium insertions lead to a real single particle potential which can be
written as, 
\begin{eqnarray} U_2(p,k_f) &=& {m_\pi^5 \over (4\pi f_\pi)^4}\bigg[ 
\int_0^{(u-x)/2}\!\! d\xi\, 8\xi+ \int_{(u-x)/2}^{(u+x)/2}\!\! d\xi
{1\over x}(u^2-(2\xi-x)^2 ) \bigg]  \bigg\{ {4\over \sqrt{1+\xi^2} } \ln(\xi+
\sqrt{1+\xi^2}) \nonumber \\ && \times  \Big[ g_A^4(11\xi^4+16\xi^2+8)-2g_A^2
(5\xi^4 +7\xi^2+2) -(1+\xi^2)^2 \Big] +(1-14g_A^2+61 g_A^4) \xi \nonumber \\ &&
+2(1+2g_A^2+5g_A^4) \xi^3 +\Big[ 6(15g_A^4-6g_A^2-1)\xi +4(11g_A^4 -10 g_A^2
-1) \xi^3 \Big] \ln{m_\pi \over 2\Lambda} \bigg\} \,. \end{eqnarray}
In order to avoid very lengthy analytical expressions we have stayed at a
one-parameter integral representation in eq.(16). 

Finally, we give the complete
expression for the power divergences specific to cut-off regularization, 
\begin{equation} U_2(p,k_f)={2\Lambda \, k_f^3 \over (4\pi f_\pi)^4 } \Big[ -10
g_A^4 M +(3g_A^2+1)(g_A^2-1) \Lambda\Big] \,. \end{equation}
The term linear in the cut-off $\Lambda$ stems from iterated $1\pi$-exchange
with a contribution of the Hartree- and Fock diagram in the ratio $4:1$. The
term quadratic in the cut-off, on the other hand, originates from irreducible
$2\pi$-exchange. Note that the momentum independent contribution to the 
two-body potential $U_2(p,
k_f)$ in eq.(17) is just twice its contribution to the energy per particle 
$\bar E(k_f)$. This relative factor of 2 is typical for a momentum independent 
NN-contact interaction, to which the power divergences are completely
equivalent, as emphasized in ref.\cite{nucmat}.  

\subsection{Results}
For the numerical evaluation of the real single particle potential $U(p,k_f)$
we use consistently the same parameters as in our previous work 
\cite{nucmat}. There, the cut-off scale $\Lambda=646.3\,$MeV has been 
fine-tuned to the binding energy per particle,  $-\bar E(k_{f0})=15.26\,$MeV. 
With this input the prediction for nuclear matter saturation density was
$\rho_0=0.178\,$fm$^{-3}$  (corresponding to a Fermi momentum of
$k_{f0}=272.7\,$MeV) and the nuclear compressibility came out as $K=255\,$MeV. 
 
In Fig.\,2, we show the total real single particle potential $U(0,k_f)$ of our
calculation for a nucleon at rest ($p=0$) as a function of the nucleon density 
$\rho =2k_f^3/3\pi^2$. The shape of the curve in Fig.\,2 is very similar to 
the nuclear matter equation of state (the so-called saturation curve). In
comparison to the energy per particle, $\bar E(k_f)$, the scale on the 
ordinate is stretched by a factor of about $3.5$. Interestingly, the potential 
depth $U(0,k_f)$ reaches its minimum close to the saturation density $\rho_0=
0.178\,$fm$^{-3}$. The actual value at that point is $U(0,k_{f0})=-53.2\,$MeV.
This value is in good agreement with the depth of the empirical optical model 
potential $U_0 \simeq -52\,$MeV deduced by extrapolation from elastic 
nucleon-nucleus scattering data \cite{phenpot,hodgson}. The average nuclear 
potential used as input in shell model calculations of nuclei has approximately
the same depth (see section 2.4 in ref.\cite{bohr}). For comparison, the 
calculation of ref.\cite{grange} based on the 
phenomenological Paris NN-potential finds a potential depth of  $U(0,k_{f0})
\simeq -64\,$MeV. In the relativistic Dirac-Brueckner approach of
ref.\cite{rolf} using the Bonn-A NN-potential a somewhat deeper real single
particle potential with $U(0,k_{f0})\simeq -80\,$MeV has been found. It should
also be noted that the potential depth  obtained in the present work results 
from a cancelation of individually large attractive and repulsive terms. For 
example, when ordering the contributions to $U(0,k_{f0})$ in third, fourth and 
fifth power of small momenta one has $U(0,k_{f0})= (-314.7+277.4-15.9)\,$MeV.
The decomposition of the same number into contributions from $1\pi$-exchange, 
iterated $1\pi$-exchange and irreducible $2\pi$-exchange reads $U(0,k_{f0})= 
(33.0-109.3+23.1)\,$MeV. Similar features hold also for the binding energy 
per particle $-\bar E(k_{f0})=15.26\,$MeV (see the discussion in section 2.5 of
ref.\cite{nucmat}).  
\bigskip

\bild{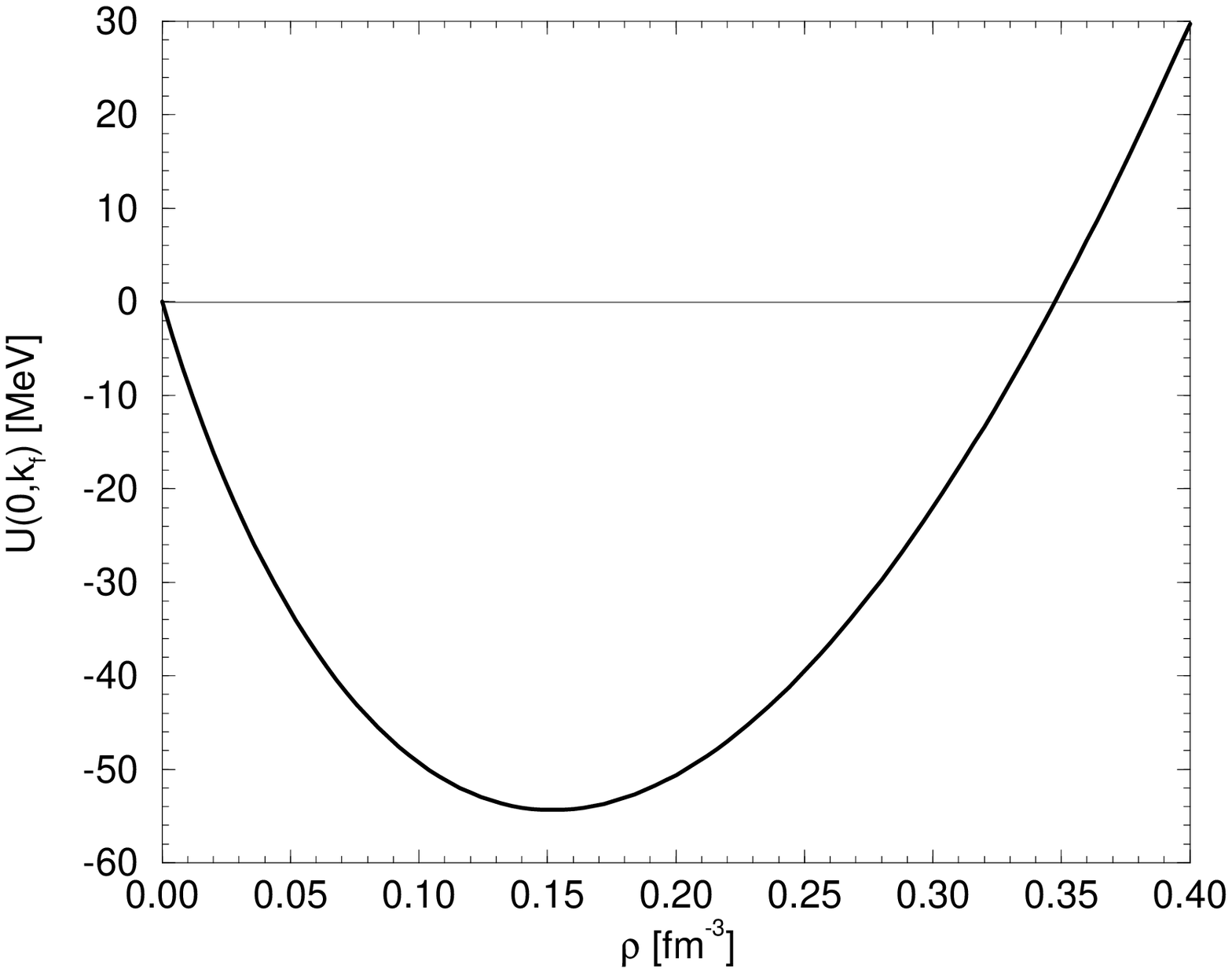}{10}
{\it Fig.\,2: The real part of the single particle potential $U(0,k_f)$ at
nucleon momentum $p=0$ versus the density $\rho=2k_f^3/3\pi^2$.}

\bigskip

We note that both the sum rule eq.(5) and the
Hugenholtz-van-Hove theorem eq.(7) hold with very high numerical accuracy in
our calculation. The validity of the Hugenholtz-van-Hove theorem eq.(7) can  
easily be proven in the present framework on the basis of the density of states
$d(\vec p_j)=(2\pi)^{-3} \, \theta(k_f-|\vec p_j|)+ \eta \,\delta^3(\vec p
-\vec p_j)$  with $| \vec p\,|=k_f$.

In Fig.\,3, the solid line shows the momentum dependence of the real 
single particle potential $U(p,k_{f0})$ at saturation density for momenta from 
zero up to the Fermi surface, $0\leq p\leq k_{f0}=272.7\,$MeV. The
dashed line in Fig.\,3 represents the total single particle energy, $T_k(p)+
U(p,k_{f0})$, i.e. the sum of single nucleon kinetic and potential energy. As
required by the Hugenholtz-van-Hove theorem the dashed line ends at the Fermi
surface $p=k_{f0}=272.7\,$MeV with the value $\bar E(k_{f0})=-15.26\,$MeV. The
total single particle energy  $T_k(p)+U(p,k_{f0})$ rises monotonically with the
nucleon momentum $p$, as it should. Note however that there is no a priori
guarantee for such a behavior in a 
perturbative calculation.  The Hugenholtz-van-Hove theorem eq.(7) determines 
also the end point of the solid line in Fig.\,3 as $U(k_{f0},k_{f0})=-54.0\,
$MeV. Since it starts with almost the same value $U(0,k_{f0})=-53.2\,$MeV at
$p=0$ the real single particle potential  $U(p,k_{f0})$ cannot be a 
monotonically rising function of $p$ in our calculation. As a matter of fact, 
one observes a downward bending of the real single particle potential
$U(p,k_{f0})$ for nucleon momenta above $p=180\,$MeV. Such a behavior is also 
found in the calculation of ref.\cite{grange}
based on the phenomenological Paris NN-potential, however, in much weaker form.
The relativistic Dirac-Brueckner calculations of ref.\cite{rolf} give, in
contrast to this, single particle potentials which rise monotonically with the 
nucleon momentum $p$.

\bigskip

\bild{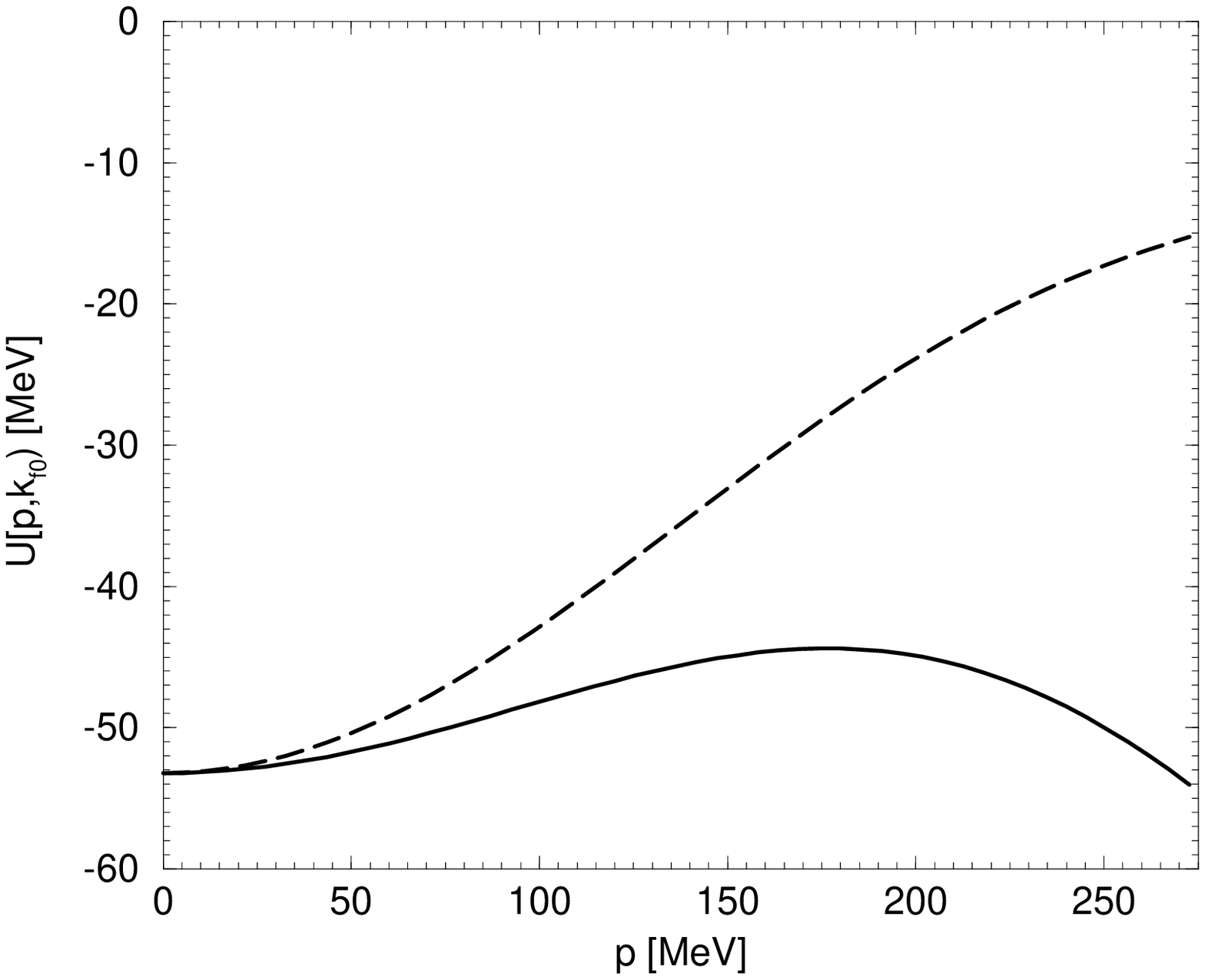}{12}
{\it Fig.\,3: The real part of the single particle potential $U(p,k_{f0})$
versus the nucleon momentum $p$ at nuclear saturation density,
$k_{f0}=272.7\,$MeV (solid line). The dashed line includes in addition the
(relativistically improved) kinetic energy $T_k(p)=p^2/2M-p^4/8M^3$.}

\bigskip

The momentum dependence of $U(p,k_{f0})$ translates into a variable 
effective nucleon mass $M^*(p)$ (the product of "$k$-mass" and "$E$-mass"
divided by the free nucleon mass, according to the nomenclature of 
ref.\cite{mahaux}) by the relation,
\begin{equation} {1 \over M^*(p)} = {1\over M} +  {1\over p} \, {\partial
U(p,k_{f0})\over \partial p} \,, \end{equation}
where $M=939\,$MeV is the (free) nucleon mass. Various many-body calculations
\cite{mahaux,grange,schuck} find that the variable effective nucleon mass
$M^*(p)$ is
reduced below $p\simeq 0.8k_{f0}$ and enhanced around the Fermi surface
$p=k_{f0}$, typically by $20\%$ on either side. The up- and downward bending of
our real single particle potential $U(p,k_{f0})$ (solid line in Fig.\,3) also 
produces such an effect, but the momentum dependence is too strong. On the
other hand, according to the sum rule eq.(5), the detailed momentum dependence 
of the components of the real single particle potential $U_{2,3}(p,k_f)$ and 
therefore also that of the variable effective nucleon mass $M^*(p)$ is of 
almost no relevance for the nuclear matter equation of state. A more suitable
quantity for comparison is the mean effective nucleon mass, 
\begin{equation} {1 \over \bar M^*} = {1\over M} +  {2\Delta U\over
k_{f0}^2}\,, \end{equation}  
with $\Delta U$ the maximal change of the potential $U(p,k_{f0})$ in the
momentum interval $0\leq p \leq k_{f0}$. In the simple case that the real 
single particle potential grows quadratically with $p$, this $\bar M^*$  
agrees with the standard definition of the effective mass, eq.(18). By 
inspection of Fig.\,3 one finds a maximal increase of our single particle 
potential $U(p,k_{f0})$ of $\Delta U=8.8\,$MeV. Via eq.(19) this shift $\Delta
U$ translates into a mean effective nucleon mass of  $\bar M^* = 0.82 M$. 
This value is compatible with empirical effective nucleon mass $M^*_{emp} = 
(0.7 -0.8) M$ derived from experimental data in the framework of 
non-relativistic shell or optical models \cite{horen,jaminon,brack}.

The real single particle potential $U(p,k_f)$ has of course a smooth 
continuation above the Fermi surface, $p > k_f$. We will not consider 
$U(p,k_f)$ in this region, for two reasons. First, most of the integral
representations for the contributions $U_{2,3}(p,k_f)$ given in section 3 do 
not hold anymore for $p > k_f$. Secondly, when going above the Fermi surface
the relevant nucleon momentum soon becomes  too large for the application of
chiral perturbation theory (which should be restricted to $p\leq 300\,$MeV), 
keeping in mind that this theory is an expansion in small external momenta 
such as  $p$ and $k_f$.  

\section{Imaginary part of the single particle potential}

In this section, we discuss the imaginary part $W(p,k_f)$ of the single 
particle potential. For the reasons just mentioned, we will restrict ourselves
to the momentum region below the Fermi surface, $0\leq p \leq k_f$. In this 
region nucleon-hole states with restricted momentum, $0\leq p \leq k_f$, can be
prepared on top of the filled Fermi sea. Their total energy is $-T_k(p)-U(p,
k_f) - i\,W(p,k_f)$, according to eq.(2). The (positive) imaginary single
particle  potential $W(p,k_f)$ accounts for the finite life time of a
nucleon-hole state 
via $\tau_{hole}^{-1}= 2W(p,k_f)$. By on-shell NN-scattering processes the
energy of a deeply bound hole-state gets redistributed among two hole-states
closer to the Fermi surface and a nucleon state in the continuum (i.e. above 
the Fermi surface). Energy and momentum conservation as well as Pauli blocking
limit the available phase space which vanishes for $p\to k_f$ as $(k_f-p)^2$
according to Luttinger's theorem \cite{luttinger}. Within the present
three-loop chiral perturbation theory calculation of nuclear matter the
contributions to the imaginary single particle potential $W(p,k_f)$ arise
entirely from iterated one-pion exchange. As done in section 3 for the real
part, we associate the contributions to $W(p,k_f)$ with the closed vacuum
diagram (see Fig.\,1) before opening of a nucleon line. Without going into
further technical details, we enumerate now the contributions to the components
$W_n(p,k_f)$ and just indicate the type of iterated $1\pi$-exchange diagram
(Hartree or Fock) and the number of medium insertions $n$. 

\noindent
i) Hartree-diagram with two medium insertions:  
\begin{eqnarray} W_2(p,k_f) &=& {\pi g_A^4Mm_\pi^4 \over (4\pi f_\pi)^4}\bigg\{
x^2 \Big( {x^2\over 5} -2u^2+{50\over 3} \Big) -3u^4 -34u^2-14 \nonumber \\ 
&&+ \Big( 9+6u^2 +{4u^3 \over x}-2x^2 \Big) \ln[1+(u+x)^2]+ \Big( 9 +6 u^2
-{4u^3 \over x}-2x^2 \Big)\nonumber \\ && \times  \ln[1+(u-x)^2]+{1\over x}(7+
15u^2-15x^2)\Big[\arctan(u+x)-\arctan(u-x)\Big]\bigg\}\,, \end{eqnarray} 
with the abbreviations $u=k_f/m_\pi$ and $x=p/m_\pi$. 

\noindent
ii) Fock-diagram with two medium insertions:  
\begin{eqnarray} W_2(p,k_f) &=& {3\pi g_A^4Mm_\pi^4\over(4\pi f_\pi)^4} \bigg\{
{u^4\over 4}+{u^2x^2\over 6}-{x^4\over 60}\nonumber \\ && -\bigg[\int_0^{(u-
x)/2}\!\!d\xi\, 4\xi+ \int_{(u-x)/2}^{(u+x)/2}\!\! d\xi {1\over 2x}(u^2-(2\xi-
x)^2 )\bigg]{1+4\xi^2 \over 1+2\xi^2}\ln(1+4\xi^2) \bigg\} \,. \end{eqnarray}  

\noindent
iii) Hartree diagram with three medium insertions:  
\begin{eqnarray} W_3(p,k_f) &=&{3\pi g_A^4M m_\pi^4 \over (4\pi f_\pi)^4}
\int_{-1}^1  \!dy\bigg\{(u^2-x^2y^2)\bigg[{2s^2+s^4\over 1+s^2}-2\ln(1+s^2)
\bigg]+{s^4 x^2(y^2-1) \over 1+s^2} \nonumber \\ &&  -{3s^4\over 2}+ 10xy \Big[
\arctan s-s +{s^3 \over 3}\Big] +(3+2u^2-2x^2y^2) \Big[s^2- \ln(1+s^2)\Big]
\nonumber \\ && + \int_0^u {2\xi^2 \over x}\,\theta(x-\xi |y|) \bigg[{2\sigma^2
+\sigma^4 \over 1+\sigma^2}-2\ln(1+\sigma^2) \bigg] \bigg\} \,. \end{eqnarray}
The auxiliary functions $s$ and $\sigma$ have been defined in eq.(12).

\noindent
iv) Fock diagram with three medium insertions:  
\begin{eqnarray} W_3(p,k_f) &=&{3g_A^4M m_\pi^4 \over (4\pi f_\pi)^4}
\int_{-1}^1\!dy  \bigg\{ \int_{-1}^1\!dz {\theta(1-y^2-z^2) \over 4\sqrt{1-y^2-
z^2}}\Big[s^2-\ln(1+s^2) \Big] \Big[\ln(1+t^2)-t^2\Big] \nonumber \\ && + 
\int_0^u \!d\xi \, {\pi \xi^2\over x} \,\theta(x-\xi |y|) \Big[\ln(1+\sigma^2)
-\sigma^2\Big] \Big( 1 - {1\over R} \Big) \bigg\}\,. \end{eqnarray}
The auxiliary functions $t$ and  $R$ have been defined in eq.(15).

\noindent
v) Hartree diagram with four medium insertions:  
\begin{eqnarray} W_4(p,k_f)&=&{6\pi g_A^4M m_\pi^4 \over (4\pi f_\pi)^4}\bigg\{
{7\over 2}-{8x^2\over 3}+{8x^4\over 15}-3 \ln(1+4x^2)+\Big(5x-{7\over 4x} \Big)
\arctan 2x \nonumber \\ && + \int_{-1}^1  \!dy\bigg[  10xy
\Big[s -{s^3 \over 3} -\arctan s\Big] +(3+2u^2-2x^2y^2) \Big[ \ln(1+s^2)-s^2
\Big] \nonumber \\ && +{3s^4\over 2}+ {s^4 x^2(1-y^2)  \over 1+s^2}
\bigg]\bigg\} \,, \end{eqnarray} 

\noindent
vi) Fock diagram with four medium insertions:  
\begin{eqnarray} W_4(p,k_f) &=&{3\pi g_A^4M m_\pi^4 \over (4\pi f_\pi)^4}
\int_0^u \!d\xi\int_{-1}^1\!dy \,{\xi^2\over x} \Big( 1 - {1\over R}\Big)  
\bigg\{\theta(x-\xi|y|)\,\theta(\xi-x)\nonumber\\ &&\times \Big[\sigma^2-
\ln(1+\sigma^2)\Big] + \theta(x-\xi)\Big[\sigma_x^2-\ln(1+\sigma_x^2)\Big]  
\bigg\} \,, \end{eqnarray}
with a new auxiliary function
\begin{equation}\sigma_x= \xi y+\sqrt{u^2-x^2+\xi^2y^2}\,.  \end{equation}

The total sum of these six contributions eqs.(20-25) evaluated at zero nucleon
momentum ($p=0$) can even be written as a  closed form expression, 
\begin{eqnarray}  W(0,k_f) &=&{3\pi g_A^4M m_\pi^4 \over (4\pi f_\pi)^4}\bigg\{
{u^4\over 2}-4u^2-{2u^2\over 1+u^2}+{\pi^2\over 12}+{\rm Li}_2(-1-u^2)\nonumber
\\ && + \Big[6+u^2+\ln(2+u^2)-{1\over 2}\ln(1+u^2) \Big] \ln(1+u^2)
\bigg\} \,, \end{eqnarray}
with $u=k_f/m_\pi$ and Li$_2(-a^{-1}) = \int_0^1 d\zeta \, (\zeta+a)^{-1} 
\ln\zeta$ denotes the conventional dilogarithmic function. Note that eq.(27) 
determines the half width of a nucleon-hole state at the bottom of the Fermi 
sea. The density dependence of $W(0,k_f)$ in eq.(27) goes approximately as  
$\rho^{4/3}$ since the first term, $u^4/2$,  in the curly bracket dominates. 
This is in fact also the only term which survives in the chiral limit
$m_\pi=0$.  

\subsection{Results}
Evidently, the results for the imaginary single particle potential $W(0,k_f)$
are completely parameterfree (to the order we are working here). At saturation
density $\rho_0=0.178\,$fm$^{-3}$ (corresponding to $k_{f0}=272.7\,$MeV) we
find from eq.(27) a half width for a nucleon-hole state at the bottom of the
Fermi sea of  $W(0,k_f)=29.7\,$MeV. This is not far from the value $W(0,k_f)
\simeq 40\,$MeV obtained in the self-consistent Brueckner calculation
of ref.\cite{grange} using the phenomenological Paris NN-potential. On the
other hand the calculation of ref.\cite{schuck} employing the Gogny D1
effective interaction finds about half of the value of ref.\cite{grange}, 
namely $W(0,k_f)\simeq 20\, $MeV. Taking the results of these two calculations 
as a reasonable benchmark our prediction for the imaginary single particle
potential at zero-momentum, $W(0,k_f)=29.7\,$MeV, can be considered as fairly
realistic.  

\bigskip

\bild{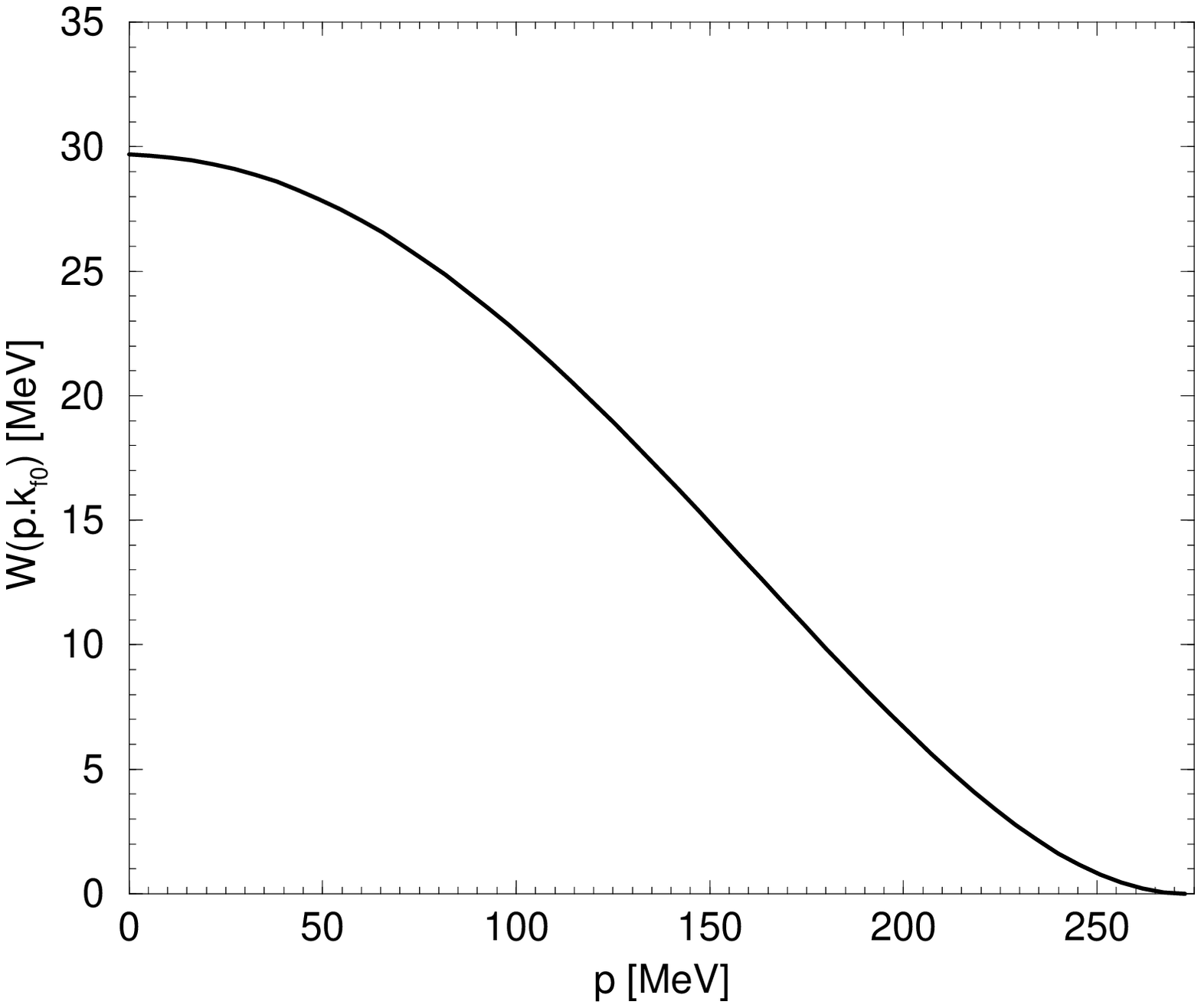}{10}
{\it Fig.\,4: The imaginary part of the single particle potential $W(p,k_{f0})$
versus the nucleon momentum $p$ at nuclear matter saturation density,
$k_{f0}=272.7\,$MeV.}  

\bigskip

In Fig.\,4, we show the momentum dependence of the imaginary part of the single
particle potential $W(p,k_{f0})$ at nuclear matter saturation density,
$k_{f0}=272.7\,$MeV. As a consequence of the decreasing phase space available  
for the redistribution of a nucleon-hole state's energy the curve in Fig.\,4 
drops with momentum $p$ and $W(p,k_{f0})$ reaches zero at the Fermi surface 
$p=k_{f0}$. From the formulas eq.(20-26) given above one can actually proof 
that both the value $W(k_f,k_f)$ and the slope $[\partial W(p,k_f)/\partial
p]_{p=k_f}$ vanish identically at the Fermi surface $p=k_f$. This is even
separately true for the contributions of the iterated $1\pi$-exchange Hartree-
and Fock-diagram.  Luttinger's theorem \cite{luttinger} is therefore satisfied
in our perturbative calculation. The imaginary single particle potential grows
quadratically near the Fermi surface,  
\begin{equation}   W(p,k_f)= C(k_f) \cdot (p-k_f)^2 + \dots \,, \end{equation}
with $C(k_f)$ a density dependent curvature coefficient. The quantity $C(k_f)$
is useful in order to characterize  the damping of the
single particle motion of valence nucleons in nuclei \cite{jeuk}. In Fig.\,5,
we show the dependence of the curvature coefficient $C(k_f)$ (given in units of
GeV$^{-1}$) on the density for $\rho \leq 0.4\,$fm$^{-3}$. One observes an
approximate $\rho^{2/3}$-dependence of $C(k_f)$. This behavior can be easily 
understood when taking the chiral limit $m_\pi=0$. In that case the
proportionality $C(k_f) \sim k_f^2$ becomes a simple consequence of (mass)
dimension counting.  

Finally, we note that the zero sum rule eq.(6) for the two-, three- and 
four-body components of the imaginary single particle potential
$W_{2,3,4}(p,k_f)$ holds with high numerical accuracy separately for the
iterated $1\pi$-exchange  Hartree- and Fock-diagram. This serves as an
important check on our calculation. 
  
\bigskip

\bild{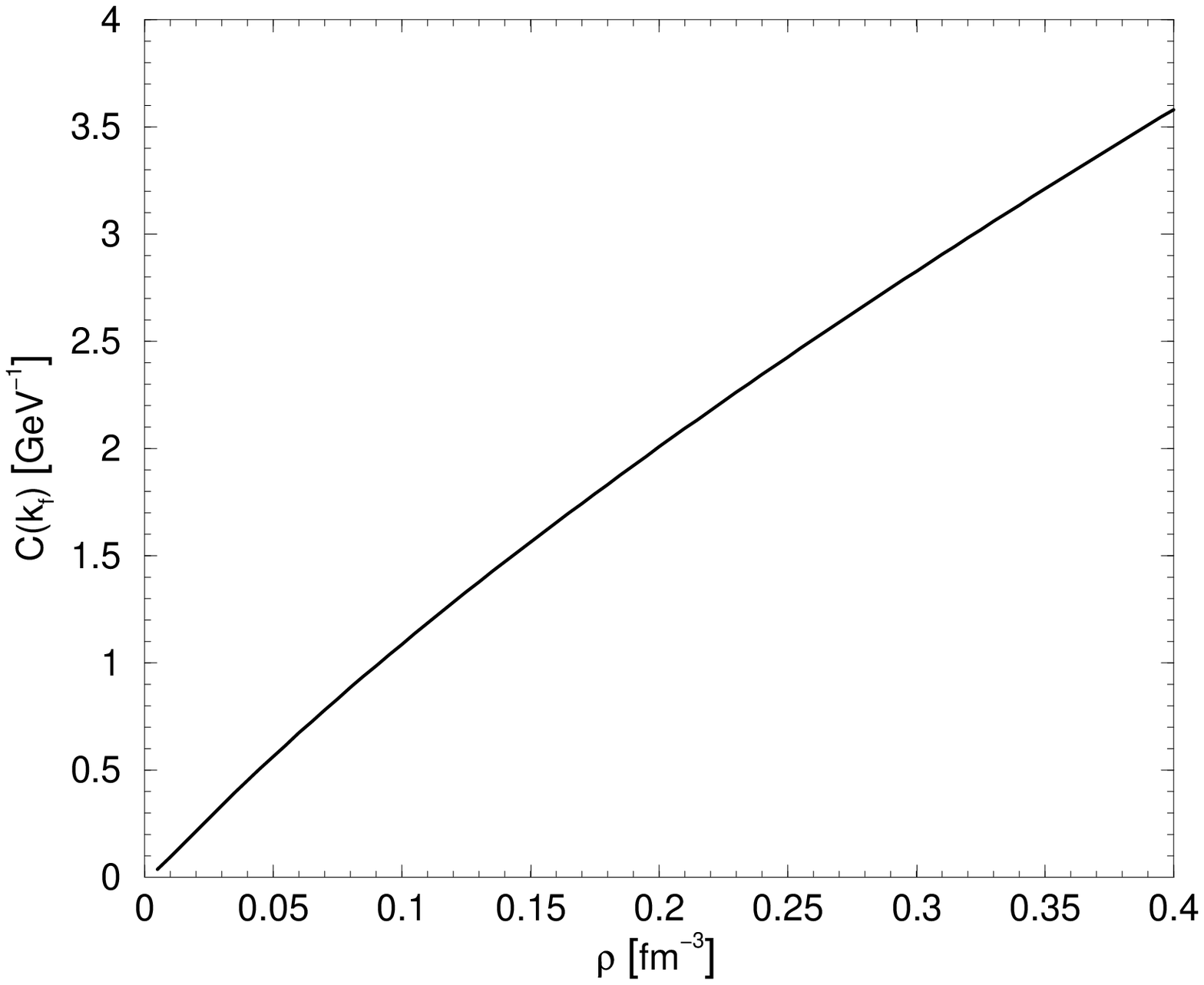}{9}
{\it Fig.\,5: The curvature coefficient $C(k_f)$ of the imaginary part of the
single particle potential $W(p,k_f)$ versus the nucleon density
$\rho=2k_f^3/3\pi^2$.}  

\section{Concluding Remarks}
In this paper, we have continued the systematic treatment of the nuclear matter
many-body problem in the framework of chiral perturbation theory by calculating
the (complex-valued) single particle potential or nuclear mean field. From the
good results of our previous paper \cite{nucmat} and the present work one can 
indeed conclude that perturbative chiral pion-nucleon dynamics is able to
produce nuclear binding and saturation (including its isospin dependence) as 
well as realistic in-medium single particle properties (including the average 
nuclear mean field). Most surprisingly, all this can be achieved with only 
one adjustable parameter, namely a single cut-off scale $\Lambda\simeq 0.65\,
$GeV which represents short distance NN-dynamics. 

Our approach to the nuclear matter problem is in many respects different from 
most other commonly used ones. First, we do not start from a so-called 
realistic NN-potential (which fits deuteron properties and NN-phase shifts), 
but we use instead  well-founded chiral pion-nucleon dynamics. From that we 
calculate analytically in-medium multi-loop diagrams up to a certain order in
a small momentum expansion. This "inward-bound" method starts from large
distances (small Fermi momentum $k_f$) and systematically generates the
pion-induced correlations between nucleons as they develop with decreasing
distance (increasing $k_f$) down to the average NN-distance in nuclear matter,
$\bar d \simeq 1.8\,$fm, which is just slightly larger than the pion Compton
wave length $m_\pi^{-1}=1.46\,$fm. 

Of course, open questions concerning the role of yet higher orders in the
chiral expansion remain and should be explored.

\end{document}